\documentclass[aps,reprint,pra,superscriptaddress,twocolumn]{revtex4-2}
\usepackage{graphicx}
\usepackage{physics}
\usepackage{geometry}
\usepackage{siunitx}
\usepackage[T1]{fontenc}
\usepackage{subfigure}
\usepackage[hidelinks]{hyperref}
\usepackage{xcolor}

\begin{document}
	\title{Probing Bandwidth and Sensitivity in Rydberg Atom Sensing via Optical Homodyne and RF Heterodyne Detection}

	\author{Dixith Manchaiah}
	\email[]{dixith.manchaiah@nist.gov}
	\affiliation{National Institute of Standards and Technology, Boulder, Colorado 80305, USA}
	\affiliation{Department of Physics, University of Colorado, Boulder, Colorado 80309, USA}

	\author{Stone Oliver}
	\affiliation{National Institute of Standards and Technology, Boulder, Colorado 80305, USA}
	\affiliation{Department of Physics, University of Colorado, Boulder, Colorado 80309, USA}
	
	\author{Samuel Berweger}
	\author{Christopher L. Holloway}
    \email[]{christopher.holloway@nist.gov}
	\author{Nikunjkumar Prajapati}
    	
	\affiliation{National Institute of Standards and Technology, Boulder, Colorado 80305, USA}

	\begin{abstract}
   Rydberg atom based sensors allow for SI traceable measurements and show promise for applications in the field of communication and radar technologies. In this article, we investigate the bandwidth and sensitivity of a Rydberg atom-based sensor in a rubidium vapor cell using Rydberg electromagnetically induced transparency~(EIT) spectroscopy. We employ a radio-frequency~(RF) heterodyne measurement technique in combination with an optical homodyne setup to extend the achievable range between sensitivity and bandwidth in a Rydberg sensor. While the bandwidth of Rydberg sensors are limited by the transit time of atoms and the Rabi frequency of the coupling field, achieving higher bandwidth through smaller beam sizes is thought to compromise sensitivity due to reduced EIT signal strength. Using optical homodyne detection, we demonstrate that sensitivity is preserved while achieving a response bandwidth of \SI{8}{\mega\hertz}. In addition, using the Rydberg sensor, we receive digital communication signals and present error vector magnitude~(EVM) measurements as a function of varying symbol rates and bandwidth of the Rydberg sensor. Furthermore, the sensor's performance is compared with a conventional RF mixer. We establish that the bandwidth of a Rydberg sensor when receiving a pure tone is not the same as the bandwidth of the sensor when receiving a modulated signal. This difference results from the spreading of symbols in the frequency domain, leading to a reduction of the signal to noise ratio~(SNR) and an accumulation of noise over the total span of the modulated signal.

    \end{abstract}
	
	\maketitle

\section{Introduction}
\label{intro}
Rydberg systems are well known for their atomic properties such as large dipole moments, polarizabilities, and long lifetimes, making them ideal test-beds for exploring various quantum technologies~\cite{Gallagher_1988,Adams_2019,Saffman_2010,Schlossberger_2024,Artusio_Glimpse_2022}. Leveraging these properties, Rydberg atoms have also emerged as prime candidates for electric field sensing, and there have been  significant efforts from the research community over the past decade~\cite{Schlossberger_2024,10972179}. 
In communications, these sensors have been demonstrated as receivers for frequency-, amplitude-, and phase-modulated signals, including transmissions such as television broadcasts~\cite{Holloway_2019,Meyer_2018,Holloway_2021,Prajapati_2022}. In the realm of metrology, they enable SI-traceable electric field measurements through energy level shifts in response to RF fields governed by Planck’s constant~\cite{sedlacek2012,Holloway_2014_broadband}, and have been demonstrated as quantum-based standards for voltage and power\cite{Holloway_2018,Holloway_2022, 10.1116/5.0090892}. Beyond these, Rydberg sensors have shown promise in radar systems\cite{Yuan_2023}, imaging\cite{Wade_2016,NoahTwoImaging2024,10.1116/5.0264378}, and atomic thermometry~\cite{Norrgard_2021,Schlossberger_2025}, illustrating their versatility as a vapor-based quantum sensing platform.

Rydberg atom-based field sensing is conventionally realized in a three-level cascade system using the phenomenon of two-photon electromagnetically induced transparency~(EIT)~\cite{PhysRevLett.98.113003,Holloway_2014,Sedlacek_2012}. New avenues such as three-photon excitation schemes~\cite{PhysRevApplied.20.L061004,Prajapati_2024}, fluorescence detection~\cite{Prajapati_2024,NoahTwoImaging2024}, and six-wave mixing~\cite{Bor_wka_2023} are being explored to improve the performance of Rydberg sensors. However, two-photon alkali systems are of prime interest due to their simple optical setup. Two key parameters for the sensors' performance are the sensitivity weak field detection and the instantaneous/response bandwidth to detect the range of frequencies which can be detected instantly. While the use of an RF local oscillator~(LO) in heterodyne configuration has been shown to significantly improve sensitivity~\cite{Simons_2019,Kumar_2017,Jing_2020}, alternative methods such as cold atoms~\cite{Tu_2024}, repumping~\cite{Prajapati_2021}, {fluorescence detection}~\cite{Prajapati_2024}, and the use of resonators~\cite{Holloway_2022_resonator} to achieve higher sensitivity are being explored. The bandwidth of Rydberg sensors is also an active area of study; a multi channel excitation scheme was used to improve the bandwidth up to $\SI{6.8}{\mega\hertz}$~\cite{Hu_2023}; and bandwidths up to $\SI{10}{\mega\hertz}$ have been demonstrated by using a six-wave mixing process~\cite{Shylla_2025,Yang_2024,Bor_wka_2023}.
    
While the sensitivity of a Rydberg sensor depends on a narrow EIT linewidth and amplitude, its bandwidth is determined by how quickly atoms decohere from the Rydberg state. This results in an inherent trade-off between sensitivity and bandwidth. Bandwidth is often reported in the literature without sensitivity, and when sensitivity is included, it is generally evaluated at a reduced beatnote frequency. Instantaneous bandwidth is typically defined by the 3dB roll-off in signal amplitude when using a pure tone input; however, the bandwidth corresponding to a modulated signal can differ due to its broader frequency content and the accumulation of noise across that range. In sensitivity or instantaneous bandwidth measurements, noise is typically estimated over a narrow frequency band around the signal of interest. In contrast, when modulated signals are used, noise accumulates over the entire frequency span, leading to discrepancies in sensor performance under realistic conditions.  Characterizing these subtleties of the sensor performance is critical for the robust deployment of Rydberg sensors in practical scenarios. Various methods have been employed to realize communication using Rydberg systems. Rydberg receivers have been demonstrated to detect a range of digital modulation formats, including binary phase-shift keying~(BPSK), quadrature phase-shift keying~(QPSK), quadrature amplitude modulation~(QAM), and on-off keying~(OOK), demonstrating their potential for advanced wireless communication applications in real-world scenarios~\cite{Holloway_2019,Song_2019,Nowosielski_2024,Wang_2025,Otto_2021,Meyer_2018,Meyer_2023,Elgee_2024}. Optimizing both the sensitivity and bandwidth of Rydberg sensors remains a crucial step toward their robust deployment in practical systems, such as digital signal reception.
	
In this work, we first present a study on the sensitivity and bandwidth of a Rydberg atom-based sensor using an RF heterodyne measurement in combination with an optical homodyne technique in $^{85}$Rubidium atomic vapor. We adopt optical homodyne technique in conjunction with RF heterodyne detection to measure sensitivity and to quantify the corresponding sensitivity bandwidth. Here, we refer to bandwidth based on the frequency at which the minimum detectable field becomes twice the value measured at low frequency which corresponds to a 3 dB drop in the signal amplitude. In the second part of the experiment, we use QPSK modulation to explore the interplay of the modulation's symbol rate and beatnote frequencies and their effects on the reception of digital communication signals using Rydberg receivers. Furthermore, we present the error vector magnitude~(EVM) measurements and also compare the atoms' demodulation to an RF mixer. 


\section{Experimental Details and EIT-AT}
\label{expt}
\begin{figure*}
		\includegraphics[width=0.9\linewidth]{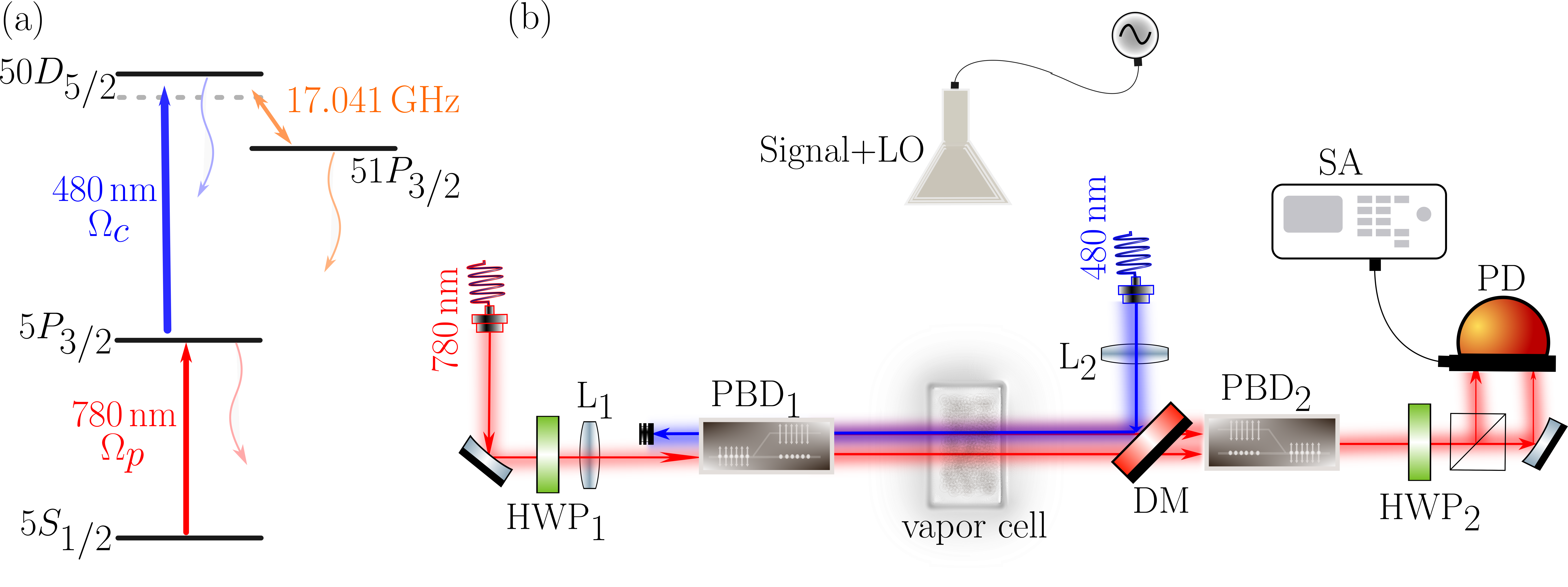}
		\caption{(a) Energy level diagram of four level system of $^{85}$Rubidium. $\Omega_c$ and $\Omega_p$ are the Rabi frequencies of coupling and probe fields. (b)Experimental schematic of the optical homodyne and RF heterodyne setup. HWP-half wave plate; PBD-polarizing beam displacer; DM-dichroic mirror; L-lens; PD-photodetector; SA-spectrum analyzer.}
		\label{fig:setup}
\end{figure*}

	The atomic system with its energy level scheme is shown in figure~(\ref{fig:setup}a). We use a typical four-level Rydberg cascade system; a $\SI{780}{\nano\meter}$ probe laser is locked to the $\ket{5S_{1/2}, F = 3}$ $\rightarrow$ $\ket{5P_{3/2}, F' = 4}$ transition using an ultra-low expansion~(ULE) cavity. A $\SI{480}{\nano\meter}$ coupling laser induces the Rydberg transition from the intermediate state to the $50D_{5/2}$ state. We then apply an RF field of $\SI{17.041}{\giga\hertz}$ to couple between the $50D_{5/2}$ and $51P_{3/2}$ Rydberg states. 
	A schematic of the experimental setup is shown in figure~(\ref{fig:setup}b). The probe laser field is derived from an external cavity diode laser~(ECDL) and is split into two orthogonal components using a polarizing beam displacer~(PBD1). In order to perform optical homodyne detection, one polarization component is used as the local oscillator~(LO), while the other serves as a signal. The coupling beam is derived from a $\SI{960}{\nano\meter}$ ECDL whose output is sent into a second harmonic generation cavity in order to double the frequency. The signal probe and the coupling beams are arranged in a counter propagating orientation in order to partially cancel the Doppler shift resulting from the motion of atoms in the vapor cell. After passing through the cell, the signal and LO probe beams are recombined into a single beam using a second PBD. Next, the combined beams pass through a half wave plate~(HWP) and polarizing beam splitter~(PBS), and are finally projected onto a balanced photodetector~(PD). The combination of signal and LO probe light in an interferometric setup with balanced detection forms a homodyne detection scheme, which serves to increase signal strength, allowing for the use of lower gain settings on a bandwidth limited photodetector~\cite{Prajapati_2022}.  The $1/e^2$ radii of the probe and coupling beams are $\SI{83}{\micro\meter}$ and $\SI{102}{\micro\meter}$, respectively, with corresponding Rabi frequencies of $\SI{18}{\mega\hertz}$ and {\SI{13}{\mega\hertz}}~(at \SI{160}{\milli\watt}). While the power of the coupling laser is varied during the measurement, the power of the signal probe beam is maintained at $\SI{5}{\micro\watt}$, and the power of the LO probe beam is maintained at $\SI{1}{\milli\watt}$. The vapor cell is $\SI{30}{\milli\meter}$ long and oriented along the beam path such that the Rayleigh ranges of the beams are well overlapped along the length of the cell. The RF signal field and RF LO field are derived from two signal generators and combined using a power combiner. The combined RF fields are then fed into a horn antenna, which is placed at a distance of $\SI{600}{\milli\meter}$ from the vapor cell.
\begin{figure}[h]
	\includegraphics[width=0.9\linewidth]{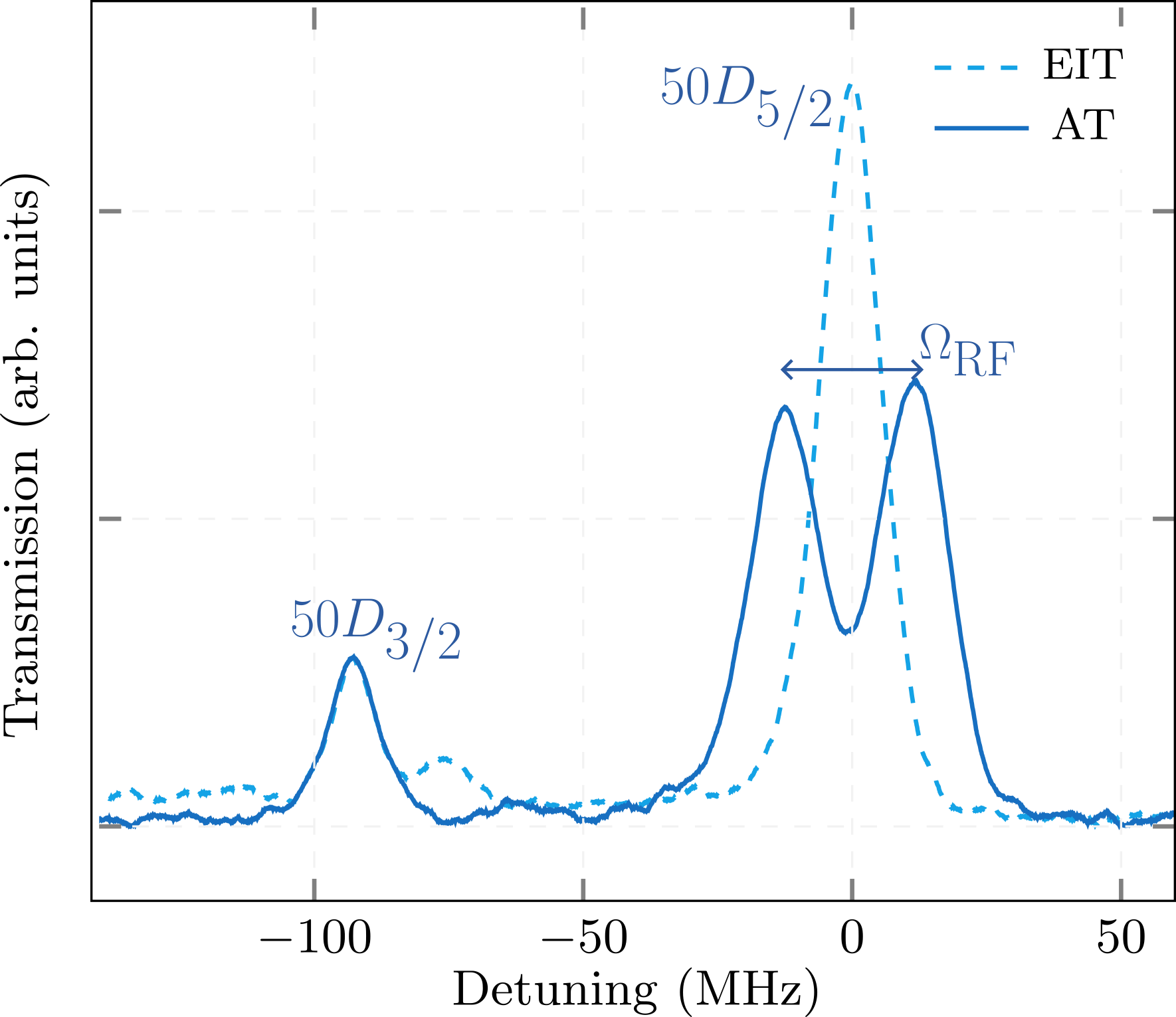}
	\caption{ Probe transmission spectrum of EIT~(dashed lines) and AT~(solid lines) as the coupling laser is scanned.} 
	\label{fig:eitcal}
\end{figure}
The probe laser is frequency-locked to the resonance using an ULE cavity, and the coupling laser is scanned across the Rydberg state, yielding an EIT peak, shown using a dashed line in figure~(\ref{fig:eitcal}). Since the coupling beam's scan range is sufficiently large, the 50D$_{3/2}$ peak is also observed. This peak is known to be $\SI{92}{\mega\hertz}$ from the 50D$_{5/2}$ peak, allowing for calibration of the frequency axis of our system. Since we are operating in the low-probe power regime, a larger photodetector gain is required, limiting the detection bandwidth. Employing optical homodyne detection provides our observed signal with an amplification factor of approximately 100, enabling the use of a lower gain setting in the photodetector. At lower gain setting, the bandwidth of the detector is higher and therefore it reduces the bandwidth limitation imposed by the photodetector. To observe Autler–Townes (AT) splitting, an RF field of $\SI{17.041}{\giga\hertz}$ is applied, driving the 50D$_{5/2}$ $\rightarrow$ 51P$_{3/2}$ Rydberg-Rydberg transition. This leads to splitting of the EIT resonance, as shown in figure~(\ref{fig:eitcal}) which is proportional to the strength of the applied RF field. Using the AT splitting, RF field inside the vapor cell is calibrated and it is discussed in appendix~\ref{appendix}.

	\section{Sensitivity and Bandwidth}
    \label{SBW}
	  We apply an RF heterodyne detection technique to determine the minimum measurable response of the atoms and the method allows us to separate the signal from the noise 
     ~\cite{Simons_2019,Jing_2020}. At this stage, both the probe and coupling lasers are frequency-locked to their respective atomic resonances using the ULE cavity to maximize the atomic response. The application of an external RF signal with the LO field induces a beating at the frequency corresponding to their difference. This beat signal manifests as a modulation in the probe transmission spectrum, which can be observed using a spectrum analyzer~(SA).
     The frequency difference between the RF LO and RF signal of interest is the beatnote frequency. This sensitivity measurement is limited by noise associated with the probe laser, photodetector, and the photon shot noise. In carrying out this measurement, the beatnote is initially set to $\SI{10}{\kilo\hertz}$. The beatnote amplitude is then maximized by optimizing the RF LO power with respect to RF signal power. The maximum beatnote amplitude is observed at RF LO  power of $\SI{-4}{dBm}$. The power of the signal field is swept from  $\SI{-40}{dBm}$ to  $\SI{-60}{dBm}$ and the resulting SNR is recorded for each RF signal power level. The SNR is used to determine the minimum detectable electric field, which defines the sensitivity $S$ of the sensor for a one-second measurement time. It is calculated using the following expression~\cite{Prajapati_2024}:
	 \begin{equation}
	S = \sqrt{(10^{(P_\text{sig}-\text{SNR})}/10) / f_\text{RBW}}\cdot C_\text{cal},
	\label{eqn:sensitivity}
	 \end{equation} 
	 where $P_\text{sig}$ is the power of the RF signal field, $f_\text{RBW}$ is the resolution bandwidth, which is set to $\SI{10}{\hertz}$ during the measurement, and $C_\text{cal}$ is the calibration factor obtained by using AT splitting. 
	 %
	 %
	  \begin{figure*}
	       \includegraphics[width = \linewidth]{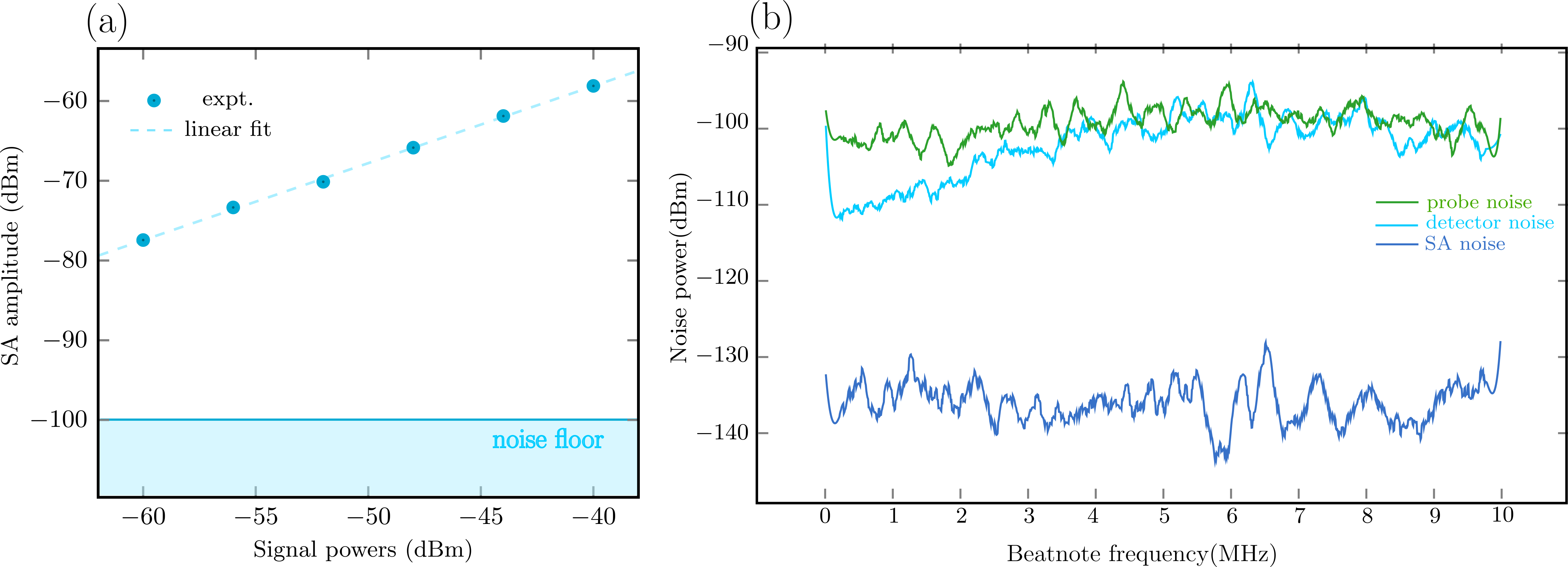}
	 	\caption{(a)RF signal field power sweep versus the SNR along with the noise floor to measure the sensitivity. (b) Plot of various noise sources present in the system.}
	 	\label{fig:sensitivity}
	 \end{figure*}
	Figure~(\ref{fig:sensitivity}a) shows a power sweep of the signal field when detuned $\SI{100}{\kilo\hertz}$ from the LO field. Each data point in figure~(\ref{fig:sensitivity}a) is an average of five traces from a spectrum analyzer, which is also used to determine the signal's SNR. The data points exhibit an increasing linear trend and are fit with a linear function. The fit is then extrapolated to the noise floor to determine the sensitivity using equation~\ref{eqn:sensitivity}. Sensitivity measurements are carried out at different beatnote frequencies, ranging from $\SI{10}{\kilo\hertz}$ to $\SI{10}{\mega\hertz}$. In the literature, bandwidth measurements are frequently presented without corresponding sensitivity measurements, and any sensitivity measurements that are provided are often measured using a lower beatnote frequency. In this experiment, sensitivities at various beatnote frequencies are recorded and the bandwidth of the Rydberg sensor is quantified based on the frequency at which the sensitivity value becomes twice the value measured at low frequency and it corresponds to a 3 dB drop in the signal amplitude. Figure~(\ref{fig:sensitivity}b) shows spectrum analyzer traces of various noise floors involved in the measurement. Each trace shown is an average of 20 traces with a resolution bandwidth of $\SI{10}{\hertz}$ in the frequency range from $\SI{10}{\kilo\hertz}$ to $\SI{10}{\mega\hertz}$. From the plot, it is evident that the system is dominated by probe laser noise, and is not limited by detector noise due to the optical homodyne technique, even when operating in the low probe power regime. 
	 
\begin{figure*}
	 	\includegraphics[width=0.9\linewidth]{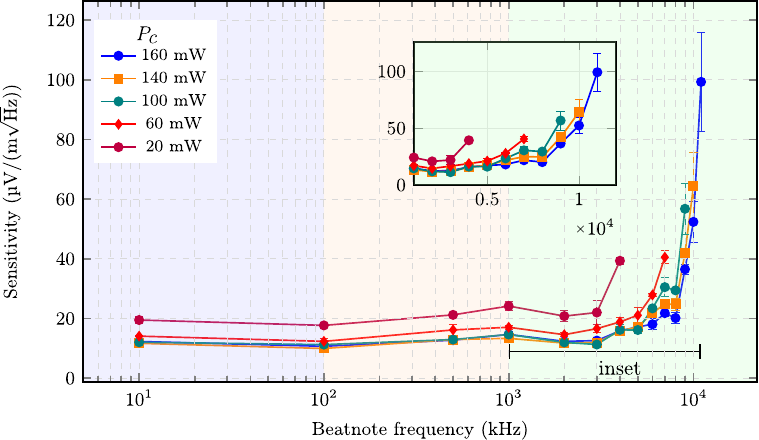}
	 	\caption{Sensitivity measured at various beatnote frequencies for different coupling powers in logarithmic scale on x-axis. Inset plot in the light green region shows the variation  from $\SI{1}{\mega\hertz}$ to $\SI{10}{\mega\hertz}$ in linear scale.}
	 	\label{fig:svsbw}
	 \end{figure*}

	 Figure~(\ref{fig:svsbw}) shows a plot of sensitivity at various beatnote frequencies for different coupling powers. It is evident from figure~(\ref{fig:svsbw}) that at a higher coupling power of $\SI{160}{\milli\watt}$, the sensitivity at $\SI{100}{\kilo\hertz}$ is $\SI{10.6}{\micro\volt\per\meter\per\sqrt\hertz}$ and is well below $\SI{20}{\micro\volt\per\meter\per\sqrt\hertz}$ up to a beatnote frequency of $\SI{8}{\mega\hertz}$. At higher beatnotes, however, sensitivity increases drastically.  The best sensitivity measured in the current setup is $\SI{9.9\pm 0.4}{\micro\volt\per\meter\per\sqrt\hertz}$ at $\SI{100}{\kilo\hertz}$ at a coupling power of $\SI{140}{\milli\watt}$. As the coupling power is decreased, sensitivity increases at lower beatnote frequencies. At $\SI{20}{\milli\watt}$ coupling power, the sensitivity is $\approx \SI{20}{\micro\volt\per\meter\per\sqrt\hertz}$ because the EIT amplitude decreases considerably as the coupling power is lowered. Each data point in figure~(\ref{fig:svsbw}) is a power sweep of the RF signal field for five traces, and sensitivity is determined as described in the last paragraph. The error bars are the deviation obtained from three repetitions. The inset plot of figure~(\ref{fig:svsbw}) shows the trend from $\SI{1}{\mega\hertz}$ to $\SI{10}{\mega\hertz}$ using a linear scale for beatnote frequencies.   
	 	 
	  Sensitivity is primarily dictated by various parameters associated with EIT spectra, such as the ratio of probe and coupling Rabi frequencies. It also depends on the noise floors involved in the system. In this measurement, the signal probe power is set to $\SI{5}{\micro\watt}$ to compensate for the Rabi frequency due to smaller beam size. However, this leads to a weaker signal which may fall below the detector noise. To achieve minimum resolvable transmission, the probe power must be maintained at a magnitude equal to the equivalent power of the noise~\cite{Prajapati_2023}. The optical homodyne technique helps in achieving the minimum resolvable power for transmission by pushing the signal above the detector noise floor. In this experiment, smaller beam sizes are also utilized to increase the Rabi frequencies, thereby decreasing the transit time. In this situation, the atoms spend less time in the vicinity of the beams, leading to collisions with other atoms or the walls of the vapor cell, causing the atoms to more rapidly decohere to the ground state. As a result, these atoms may now interact with the beams again in a shorter amount of time, ultimately yielding a higher bandwidth. The underlying physics involves the narrow EIT linewidth with longer coherence time leads to improved sensitivity which limits the atoms response time  for rapid field changes. Conversely, to increase bandwidth by reducing the coherence time often broadens the EIT linewidth, compromising sensitivity. This limitation can be mitigated by employing the optical homodyne technique along with reduced beam sizes, as discussed earlier.  At a temperature of $\SI{55}{\degreeCelsius}$, the average velocity of a Rubidium atom is $ v \approx  \SI{280}{\meter\per\second}$ and the corresponding transit time in the system is given as, 
	  
	  \begin{equation}
	  	\tau_{t} = \frac{2\times\omega}{v} = \SI{0.59}{\micro\second},
	  \end{equation}  
	which corresponds to transit rate of $\SI{1.69}{\mega\hertz}$, where $\tau_t$ is the transit time, and $\omega$ is the probe beam waist~($\SI{83}{\micro\meter}$). In conclusion, producing a higher coupling Rabi frequency through the use of smaller beam sizes and also using the optical homodyne technique enables the simultaneous improvement of both the sensitivity and bandwidth of a Rydberg sensor.
 	 
	 \section{Demodulation}
     \label{demod}
	  \begin{figure}[h]
	 	\centering
	 	\includegraphics[width=1.1\linewidth]{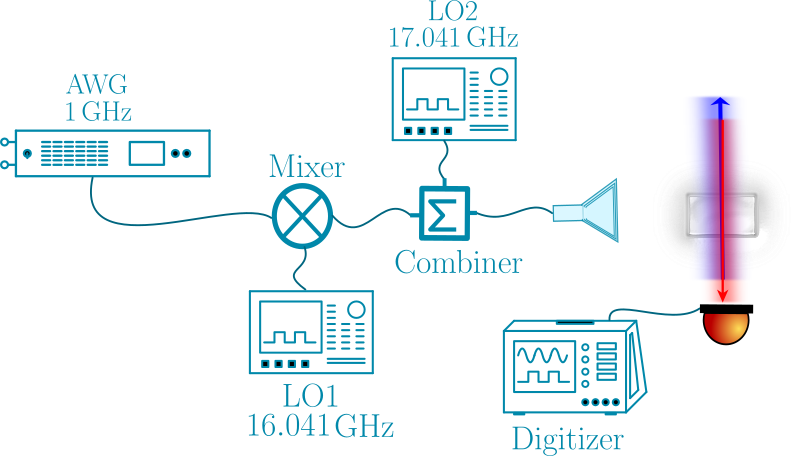}
	 	\caption{Schematic setup for the generation and readout of modulated signals using arbitrary waveform generator~(AWG).  }
	 	\label{fig:scheme}
	 \end{figure}
	 \begin{figure*}
	 	\centering
	 	\includegraphics[width=\linewidth]{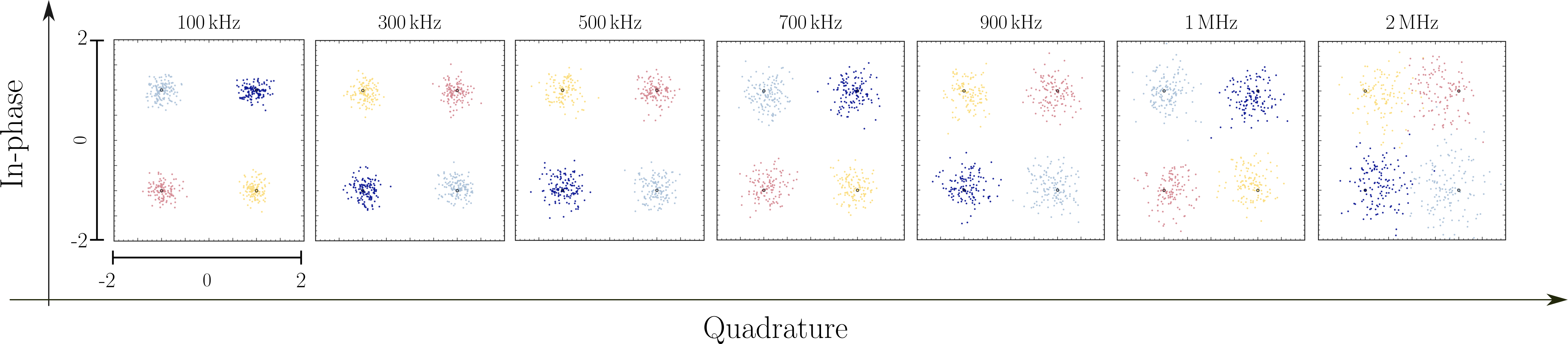}
	 	\caption{IQ constellation diagram of a QPSK modulation with varying symbol rate Each color corresponds to a unique symbol, and the black circled points correspond to the nominal position for that symbol. The beatnote frequency is maintained at $\SI{2}{\mega\hertz}$.}
	 	\label{fig:iqs}
	 \end{figure*}
	 In the second part of the experiment, we use a Rydberg sensor to receive modulated signals via RF heterodyne detection. We study the atoms' response to different beatnote frequencies and symbol rates (the symbol rate is defined as the number of transmitted symbols per second contained in a modulated signal). To produce and receive the signal, we first produce a $\SI{1}{\giga\hertz}$ carrier wave using a Keysight M8190A~\cite{Note} arbitrary waveform generator~(AWG). The carrier signal is mixed with a $\SI{16.041}{\giga\hertz}$ RF LO1, yielding a $\SI{17.041}{\giga\hertz}$ RF signal. Using a power-combiner, this RF signal is then combined with  LO2 and sent into free space using an RF horn antenna, which is pointed towards the vapor cell. The frequency of LO2 is detuned from the frequency of the RF signal to produce the desired beatnote. In this RF heterodyne scheme, the atoms act as a frequency down-mixer; the beatnote containing the signal modulation is encoded in the transmission signal of the probe laser used for Rydberg EIT. After collecting the probe transmission via photodetector, the detector's analog output is converted to a digital signal via a Keysight M9734G digitizer~\cite{Note} (note that the digitizer and waveform generator must be synchronized on the same reference clock). A schematic of the setup is shown in figure~(\ref{fig:scheme}). For each measurement, we take the average of five traces from the digitizer.

	  \begin{figure}[h]
	  	\includegraphics[width=\linewidth]{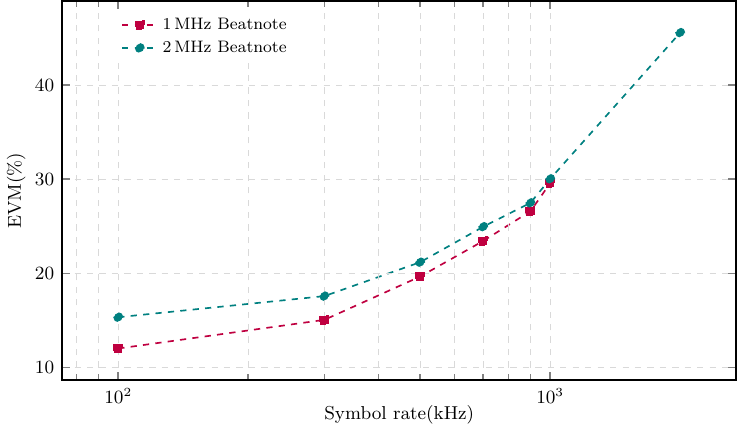}
	  	\caption{Plot of EVM versus symbol rates at beatnote frequencies of $\SI{1}{\mega\hertz}$ and  $\SI{2}{\mega\hertz}$.}
	  	\label{fig:evmsymbol}
	  \end{figure}
      
\begin{figure}[h]
	  	\subfigure[]{\includegraphics[width=\linewidth]{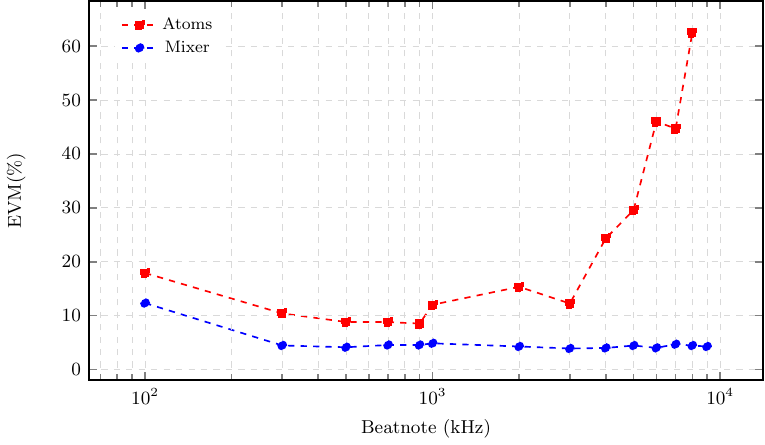}}
	  	\subfigure[]{\includegraphics[width=\linewidth]{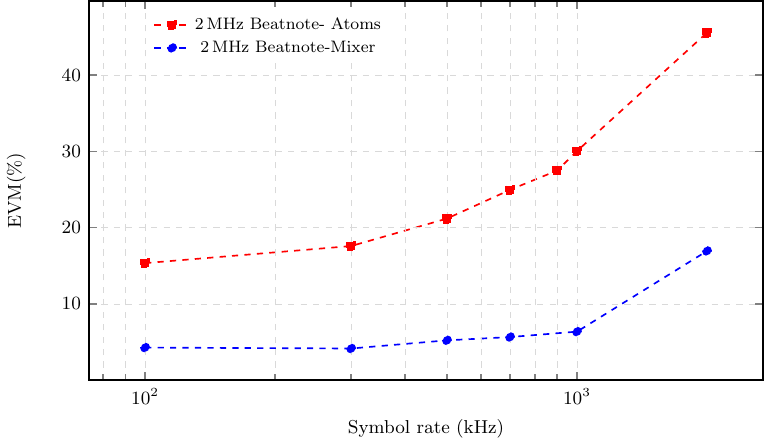}}
	  	\caption{(a) Comparison of behavior of atoms demodulation with respect to RF mixer at various beatnote frequencies at $\SI{100}{\kilo\hertz}$ symbol rate. (b) Comparison of atoms demodulation with respect to RF mixer at various symbol rates at $\SI{2}{\mega\hertz}$ beatnote.}
	  	\label{fig:evmbn}
	  \end{figure}

     For this work, we apply a quadrature-phase-shift-key~(QPSK) modulation to the carrier signal, where each modulation is comprised of 511 transmitted symbols. The data collected from the digitizer is mixed from the beatnote frequency down to baseband. Next, it is Fourier transformed and filtered using a low-pass filter set to a cutoff frequency of ${2.1}$ times the symbol rate. The signal is then inverse Fourier transformed back to the time domain. Taking either the real or imaginary part of the result, one obtains the quadrature~(Q) or in-phase~(I) components of the signal, respectively. We then compare each transmitted symbol (i.e. each data point's measured position in the complex plane) to its nominal counterpart and calculate error vector magnitude~(EVM). EVM is calculated by taking the RMS amplitude of the error vectors of all transmitted symbols. Figure~(\ref{fig:iqs}) shows IQ plots for varying symbol rates. As the symbol rates are increased, the spread in the time domain signal increases, reducing the total SNR. The reduction in SNR manifests as spread in IQ plot as shown in the figure~(\ref{fig:iqs}). Figure~(\ref{fig:evmsymbol}) plots EVM as a function of symbol rate with a fixed beatnote frequency and shows the increase in EVM with increasing symbol rate at beatnote frequencies of $\SI{1}{\mega\hertz}$ and $\SI{2}{\mega\hertz}$. Furthermore, in order to benchmark the quality of the atom response, we also compare these results with the output of a conventional RF mixer. Figure (\ref{fig:evmbn}a) plots EVM versus beatnote at a fixed symbol rate of $\SI{100}{\kilo\hertz}$ for both the atoms and a conventional mixer. At lower beatnote frequencies, higher EVM is observed due to $1/f$ noise in the system. As the beatnote is increased, both the atoms and the conventional RF mixer show a flat response up to approximately $\SI{3}{\mega\hertz}$ and beyond this point, the atoms display a rapidly increasing EVM. 

     Moving to figure (\ref{fig:evmbn}b), which plots EVM versus symbol rate at a fixed beatnote of $\SI{2}{\mega\hertz}$ for both systems, we again see the conventional mixer displaying a generally flat response (although, the EVM does begin to increase at higher symbol rates as we approach the bandwidth limitation of the mixer) while the atoms instead show an increasing EVM as symbol rate is increased. Based on the above measurements, atom's EVM increases with increasing symbol rates  and also increasing beatnote frequencies. This increase in EVM can be attributed to various noise sources present in the system such as probe laser noise and photodetector noise. During the sensitivity bandwidth measurement in the first part of the experiment with a pure tone, noise level is chosen around the beatnote frequency, which is of narrow range, whereas when a signal has QPSK modulation the noise present in the entire frequency span contribute to the system leading to increase in EVM even at lower atomic bandwidth. Based on this,
     we arrive at the conclusion that the bandwidth of the Rydberg sensor due to a pure tone is not the same as the bandwidth when a modulated signal is present. This is also evident from the fact that, in frequency space, the symbols are spread across the total signal, reducing the overall SNR. The result is an increase in the EVM as symbol rates and beatnotes are increased~\cite{Nowosielski_2024, Holloway_2019}. Furthermore, phase fluctuations during bit transitions, intermodulation effects, and system nonlinearities can increase the EVM, which in turn reduces the performance of the atomic sensor. In spite of these limiting factors, digital communication is viable using Rydberg receivers. To improve upon the system presented in this work, one could include multiple narrow band communication channels or utilize an array of spatially distributed probe fields~\cite{Otto_2021,Meyer_2018}.

\section{Conclusion}
\label{concl}
In summary, we have demonstrated that the achievable range of both sensitivity and bandwidth in a Rydberg atom-based sensor can be extended by employing optical homodyne and RF heterodyne techniques. The sensitivity bandwidth of about $\approx \SI{8}{\mega\hertz}$ is obtained by maintaining the smaller beam sizes which leads to increased transit time contribution, meanwhile sensitivity is maintained below $\approx \SI{20}{\micro\volt\per\meter\per\sqrt\hertz}$ throughout by adapting optical homodyne method. The best achieved sensitivity in the current setup was $\SI{9.9}{\micro\volt\per\meter\per\sqrt\hertz}$. Additionally, we utilized the system to receive digital communication signals, primarily QPSK. We characterized the atomic demodulation performance across various symbol rates and beatnote frequencies by analyzing EVM measurements. A comparative study between conventional RF mixer demodulation and atomic demodulation was presented by varying beatnote frequencies and symbol rates. Furthermore, we demonstrated that the bandwidth of the Rydberg sensor measured with a pure tone differs compared to bandwidth while receiving a modulated signal. This work also demonstrates the viability of Rydberg sensors as a promising platform for advanced communication systems and other applications, including radar and sensing technologies.  

\section*{Acknowledgments}
	This research was developed with funding from NIST. A contribution of
the U.S. government, this work is not subject to copyright
in the United States. 

\section*{Conflict of Interest}
The authors have no conflicts to disclose.

\section*{Data Availability Statement}
All of the data presented in this paper is available at https://doi.org/10.18434/mds2-3975

\section*{Disclaimer}
Certain commercial equipment, instruments, software, or materials, commercial or non-commercial, are identified in this paper in order to specify the experimental procedure adequately. Such identification is not intended to imply recommendation or endorsement of any product or service by NIST, nor does it imply
that the materials or equipment identified are necessarily the best available for the purpose.

\appendix
\section{Calibration}
\label{appendix}
The strength of the field can be obtained by using the equation,

\begin{equation}
    \abs{E} = \frac{\hbar\Omega_\mathrm{RF}}{d_\mathrm{RF}},
    \label{efield}
\end{equation}
where $E$ is the magnitude of the applied RF field, $d_\mathrm{RF}$ is the dipole moment of the RF transition and  $\Omega_\mathrm{RF}$ is the Rabi frequency of the RF transition. Further, splitting is used to calibrate the system and correct for changes in the electric field caused by the vapor cell, allowing accurate measurement of the field inside the cell. To calibrate the RF field inside the vapor cell, power of RF field is varied from $\SI{0}{dBm}$ to $\SI{7}{dBm}$ and the corresponding AT splittings are measured and the magnitude of electric field is obtained using the equation~(\ref{efield}).  
\begin{figure}[h]
	\includegraphics[width=0.8\linewidth]{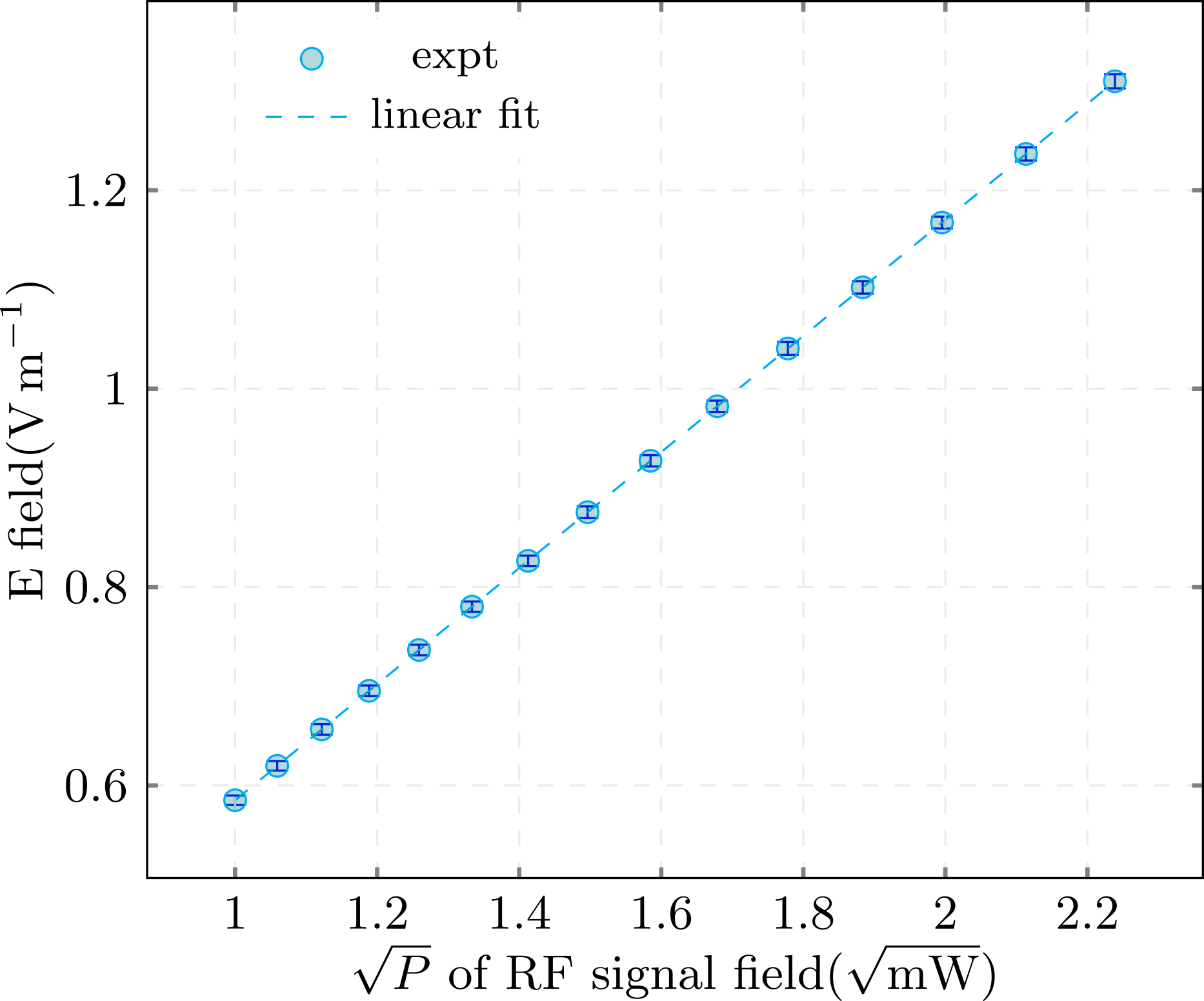}
	\caption{Calibration of RF horn antenna. $\sqrt{P}$ is the power of applied RF field.  }
	\label{fig:cal}
\end{figure}
Figure~(\ref{fig:cal}) shows the plot of electric field versus the square root of the power of RF field. The electric field at the vapor cell location can be accurately determined at higher RF power levels and subsequently extrapolated to estimate the field strength at lower output powers of the signal generator. Each point in the plot is obtained by taking an average of five traces of an AT splitting spectrum and then fitting the peaks to a double Gaussian curve to obtain the frequency separation. The error bars on the data points are the fit errors with respect to the spectrum. In figure~(\ref{fig:cal}), the data points show a linear trend. The dashed line is a linear fit to the data points whose slope is the calibration factor  $C_\mathrm{cal} = 0.58$, which is essential for measuring the sensitivity of the Rydberg sensor~\cite{Robinson_2021,Anderson_2014}.  Once the applied field is calibrated, the RF LO is introduced from a different signal generator summed using a power combiner along with the signal field. While calibrating the field at the vapor cell radiating through horn antenna, all the RF components were included but LO field was set to off.

\bibliography{reference}	

\begin{thebibliography}{46}
\bibitem [{Not()}]{Note}%
  \BibitemOpen
  \href@noop {} {}\bibinfo {note} {Certain commercial equipment, instruments, software, or materials, commercial or non-commercial, are identified in this paper in order to specify the experimental procedure adequately. Such identification is not intended to imply recommendation or endorsement of any product or service by NIST, nor does it imply that the materials or equipment identified are necessarily the best available for the purpose.}\BibitemShut {Stop}%
\end{thebibliography}
\end{document}